\documentstyle[11pt,aaspp4]{article}
\received{00 March 1996}
\accepted{00 1996}
\slugcomment{To appear in Ap. J. Letters}
\lefthead{Yusef-Zadeh, Roberts \& Wardle}
\righthead{High Velocity Ionized Gas}

\def\kms    {\hbox{km{\hskip0.1em}s$^{-1}$}}    
\def\msol   {\hbox{$M_\odot$}}                  
\def\dasec  {\hbox{$.\!\!^{\prime\prime}$}}     

\begin{document}

\title{Anomalous Motion of Ionized Gas \\
in the Sickle (G0.18--0.04) Near the Galactic Center}

\author{F. Yusef-Zadeh}
\affil{Department of Physics and Astronomy, Northwestern University, 
Evanston, Il. 60208 (zadeh@ossenu.astro.nwu.edu)}

\author{D. A. Roberts}
\affil{NCSA, 405 N. Mathews Ave, Urbana, IL 61801
 (droberts@lai.ncsa.uiuc.edu)}
\author{M. Wardle}
\affil{Special Research Centre for Theoretical Astrophysics,
University of Sydney, NSW 2006, Australia
(wardle@physics.usyd.edu.au) }

\begin{abstract}

We present VLA measurements of H92$\alpha$ radio recombination line
emission from an unusual HII region G0.18--0.04, the ``Sickle'', with
spatial and spectral resolutions of 27\dasec8 $\times$ 24\dasec9 and 14
\kms, respectively.  These observations detected two new kinematic
components of ionized gas whose velocities differ greatly from the +25
\kms\ molecular cloud surrounding the Sickle.  One component is highly
red-shifted with peak velocity of about +150 \kms\ and the other is a
blue-shifted velocity feature peaking near $-35$ \kms.  Neither of these
high-velocity features have molecular counterparts.  The blue-shifted
feature is forbidden in the sense of Galactic rotation and coincides with
the prominent nonthermal filaments crossing the Sickle, thus suggesting
that they are physically associated with each other.  The results presented
here are interpreted in terms of ionized gas being accelerated away from
the surface of the cloud associated with the Sickle region, either by the
magnetic field associated with the nonthermal filaments or by the stellar
winds from the hot helium stars near G0.18--0.04.   
\end{abstract}


\vfill\eject

\section{Introduction}

The unusual source G0.18--0.04, the ``Sickle'', is where the nonthermal
filaments (NTFs) of the Galactic center Arc appear to coincide with a
thermal source at the Galactic plane (Yusef-Zadeh, Morris \& Chance 1984;
Yusef-Zadeh \& Morris 1987a).  The filaments are long, narrow
synchrotron-emitting features tracing organized magnetic fields which run
perpendicular to the Galactic plane (Yusef-Zadeh \& Morris 1987b).
G0.18--0.04 is one of the most interesting regions in the Galactic center
because of its potential to provide important clues as to the nature of the
acceleration of relativistic particles of the NTFs and on the ionization
mechanism of thermal gas.

Radio recombination and molecular line studies of G0.18--0.04 detected 
ionized thermal gas with a Sickle-shaped appearance at a velocity near 
+40 \kms\ delineating the eastern edge of the +25 \kms\ molecular 
cloud (Yusef-Zadeh, Morris, \& van Gorkom 1989; Serabyn \& G\"usten 
1991).  G0.18--0.04 is thought to be photoionized by a number of hot 
helium stars that have been discovered nearby (Moneti, Glass \& 
Moorwood 1991; Figer, McLean \& Morris 1995; Cotera et al 1996), and 
the infrared and radio recombination line ratios are consistent with 
this picture (Harris et al 1994; Timmermann et al 1996; Lang, Goss \& 
Wood 1997; Simpson et al. 1997)

Interferometric CS observations by Serabyn \& Morris (1994) have shown 
clumps of molecular gas aligned along the NTFs, supporting the 
suggestion that the NTFs arise through an interaction with a Galactic 
Center molecular cloud with the strong, ambient magnetic field 
believed to permeate the region (Serabyn \& G\"usten 1991).  In this 
scenario, the interaction between the magnetic field and the ionized 
gas at the cloud surface loads energetic particles onto the field 
lines and the resultant synchrotron emission produces the NTF.

In this {\it Letter} we present low spatial and spectral resolution 
(26$^{\prime\prime}$ and 14 \kms) observations of G0.18--0.04.  The 
observations complement the detailed high-resolution 
(6$^{\prime\prime}$ and 8.5 \kms) H92$\alpha$ observations of Lang et 
al (1997) by their increased sensitivity to low surface brightness 
emission.  We report the detection of two extreme low- and 
high-velocity ionized features in the Sickle at $V_{LSR} = -35$ and 
+150 \kms.\ The low-velocity features are associated with the sites of 
interaction of the NTFs with the Sickle, and further strengthen the
connection between the Sickle and the NTFs.

\section{Observations}

H92$\alpha$ observations of the Sickle were carried out on 14 July 
1988 using the D-configuration of the Very Large Array of the National 
Radio Astronomy Observatory\footnote{The National Radio Astronomy 
Observatory is a facility of the National Science Foundation, operated 
under a cooperative agreement by Associated Universities, Inc.}.  A 
preliminary account of this observation was described by Yusef-Zadeh 
et al (1989).  This observation was centered at $\rm 
\alpha(1950)=17^h43^m 05^s, \delta(1950) = -28^\circ 48^\prime 
45^{\prime\prime}$.

NRAO 530 and 3C 48 were used as phase and flux density calibrators.  
Bandpass solutions were obtained using both a one hour observation of 
3C 84 and the periodic observations of the complex gain calibrator, 
NRAO 530.  The solutions obtained using NRAO 530 were superior to 
those obtained using 3C 84, due to the fact that they were determined 
periodically (every 30 min) in time and could track the short time 
scale variation in the bandpass; thus, the bandpass was corrected 
using NRAO 530.  The correlator was set to observe 32 channels in 
right circular polarization with a total bandwidth of 12.5 MHz 
centered on $V_{LSR} = +60$ \kms.  After on-line Hanning smoothing, 
the data covered a velocity range between $-162 < V_{LSR} < +275$ 
\kms\ with a channel resolution and separation of 14 \kms.  After 
careful editing of short-spacing visibilities, standard calibration 
was carried out.  In order to emphasize the weak, extended structures, 
the visibility data were naturally weighted and tapered at 5 
k$\lambda$ giving an angular resolution of 27\dasec8 $\times$ 
24\dasec9 (PA=61$^\circ$).  The continuum channels were fitted and 
subtracted in the visibility domain using UVLIN in AIPS; the resulting 
continuum-subtracted data set was imaged.  The continuum image was 
formed by averaging the visibility data in the line-free channels.  
The typical rms noises for a single line channel and for the continuum 
are $\approx$ 0.55 and 1.58 mJy beam$^{-1}$, respectively.  The 
negative features near bright sources in the final images are a result 
of structure in spatial frequencies smaller than those sampled in 
these observations.  The images of integrated line emission and 
velocity fields were created with the MOMENT program in the MIRIAD 
software package of the Berkeley-Illinois-Maryland-Association (BIMA).  
During the moment analysis, the line intensity was used only where the 
emission was above 2 mJy beam$^{-1}$ (signal-to-noise ratio [S/N] 
$\approx$ 4).

\section{Results}  

The bottom right panel of Figure 1 shows gray scale and contour 
representations of the continuum emission from the inner 20 pc of the 
Sickle feature.  The continuum contours show the diagonal SE-NW linear 
feature running perpendicular to the Galactic plane and crossing the 
Sickle.  In high-resolution observations (Yusef-Zadeh \& Morris 
1987a,b), the linear feature is resolved into a system of narrow and 
long NTFs.  This low-resolution continuum image shows clearly that the 
system of linear filaments become rather discontinuous 
and weaker in surface brightness as they cross the Galactic 
plane and extend to the northwest of the Sickle.  The continuum 
emission associated with the linear feature peaks at a flux density of 
$\approx$ 150 mJy near $\rm \alpha(1950)=17^h 43^m 10^s, \delta(1950) 
= -28^\circ 49^\prime$.  In high-resolution images, this peak feature 
appears to be extended to the northern half of the Sickle; this 
feature is called the ``Wake'' in the schematic diagram of Yusef-Zadeh 
\& Morris (1987a).  The circular-shaped source G0.15-0.05, the 
``Pistol'', is the brightest continuum feature in Fig.  1 located near 
$\rm \alpha(1950)=17^h 43^m 05^s, \delta(1950) = -28^\circ 49^\prime 
$.  A north-south feature near the southern half of the Sickle at $\rm 
\alpha(1950)=17^h 42^m 55^s, \delta(1950) = -28^\circ 50^\prime 02''$
 in extent is also noted.  The three panels surrounding the 
continuum image of Fig.  1 show three spectra taken toward positions 
marked as crosses on the continuum image.

The most interesting result is the detection of extended blue-shifted 
ionized gas which is forbidden in the sense of the Galactic rotation.  
Figure 2 shows a gray-scale image of the H92$\alpha$ line emission, 
integrated between $-63$ and $-7$ \kms, overlayed with contours of 
continuum emission.  A typical spectrum of this new feature is shown 
in position 1 of Fig.  1 with a S/N of $\approx$ 7.6 and a peak flux 
density of 4.2 mJy at $\approx -36.0$ \kms.  Most of the blue-shifted 
velocity feature is distributed in the diffuse region to the east of 
the Sickle and to the north of the nonthermal linear feature.

The integrated line emission peaks at $\rm \alpha(1950)=17^h 43^m 
3.\!\!^s9, \delta(1950) = -28^\circ 47^\prime 55^{\prime\prime} $, 
coincident with the location of one of the NTFs having the continuum 
flux density of 83.5 mJy beam$^{-1}$ as it crosses the Sickle.  The 
brightness of the continuum emission from the diagonal linear feature 
becomes rather weak exactly where the blue-shifted ionized feature 
peaks.  High-resolution radio continuum images of this area are 
dominated by the narrow, nonthermal filaments (Yusef-Zadeh \& Morris 
1987a).  Assuming that the emitting gas for this particular 
component is in LTE with the 
line-to-continuum ratio of 5\% and an electron temperature of 9800 K 
(see below), the total ionized mass and the average electron density 
for the negative velocity feature 
are estimated to be $\approx$30\msol\ and 150 cm$^{-3}$, respectively.
These values are based upon assumed model geometries and are uncertain 
by a factor of three.

The other peaks in integrated emission at $\rm \alpha(1950)=17^h 43^m 
3.\!\!^s 5, \delta(1950) = -28^\circ 46^\prime 45^{\prime\prime}$ and $\rm 
\alpha(1950)=17^h 42^m 58^s, \delta(1950) = -28^\circ 48^\prime 
15^{\prime\prime}$ are generally consistent with the high-resolution 
observations of Lang et al (1997), who noted blue-shifted velocity 
features in the region where the $-35$ \kms\ feature crosses the 
northern half of the Sickle (see their L2 and L3 spectra) and at the 
southern tip of the Sickle (their L8 line profiles).  Blue-shifted 
emission is present in the [NeII] 12.8 \( \mu \)m spectrum taken 
toward the Sickle (see Fig 6 of Serabyn \& G\"usten 1991), and 
possibly in the [OIII] 88 \( \mu \)m spectrum taken by Timmermann et 
al (1996) towards the southernmost peak of the integrated line 
emission in Fig.  2 (see their Fig.  3a).

Figure 3 is a gray-scale representation of the highest red-shifted 
velocity features between +106 and +205 \kms\ with continuum emission 
contours superposed.  Two highly red-shifted components are noted in 
this figure.  One is the velocity feature associated with the Pistol 
located to the south of the linear feature having a peak velocity of 
+125 \kms (Yusef-Zadeh et al 1989; Lang et al 1997).  The second 
component is a new high-velocity redshifted feature distributed close 
to the peak continuum emission from the linear feature.  This feature 
has a peak velocity of $\approx 150$ \kms\ at $\rm \alpha(1950)=17^h 
43^m 10.\!\!^s 3, \delta(1950) = -28^\circ 48^\prime 27^{\prime\prime} 
$.  Position 2 of Fig.  1 presents the spectrum of the peak line 
emission with a flux density of 5.7 mJy beam$^{-1}$ and a 
corresponding continuum flux density of 100 mJy beam$^{-1}$.  This 
newly-detected ionized feature has the highest radial velocity in the 
Galactic center region with the exception of Sgr A West.  The total 
ionized mass and the average electron density for this velocity 
component are estimated to be $\approx$12\msol\ and 170 cm$^{-3}$, 
respectively.

Figure 4 shows the velocity distribution of ionized gas ranging 
between $-100$ and +190 with contours of total intensity superposed.  
Note that the extent of high-velocity red-shifted gas beyond +90 \kms\ 
is not limited to the Pistol but also to the region to the southern 
half of the Sickle.  The new velocity feature peaks at $\rm 
\alpha(1950)=17^h 42^m 55.3^s, \delta(1950) = -28^\circ 49^\prime 
33^{\prime\prime}$.  The spectrum of this +92 \kms\ velocity feature 
with the peak line emission of 7 mJy is observed in Position 3 of Fig.  
1.  The total ionized mass and electron density are estimated to be 
similar to the +150 \kms velocity feature.

An accurate estimate of the thermal continuum emission from these 
diffuse ionized features is quite difficult to make in the presence of 
the non-thermal emission from the NTFs.  However, if the ionized gas 
is assumed to be LTE and that the abundance of singly ionized helium 
relative to singly ionized hydrogen (Y$^+$) is 10\%, the {\em upper 
limits} to the electron temperature can be estimated.  For positions 
1, 2, and 3 (see Fig.  1) upper limits to the electron temperatures 
are estimated to be 9800, 6100, and 2900 K, respectively.  The largest 
uncertainty from these estimates comes from the fact that the thermal 
continuum flux cannot be distinguished from nonthermal continuum.  
Unlike positions 1 and 2, which are near NTFs, position 3 (the 
southern half of the Sickle) does not appear to be contaminated by any 
emission from NTFs, thus the estimated electron temperature at this 
position is not an upper limit, but rather is probably the actual 
value.  This low value of electron temperature is less than that of 
ionized gas ranging between 4600 K and 7000 K in Sgr A West (Roberts 
and Goss 1993; Yusef-Zadeh, Zhao and Goss 1995).  It should be stated 
that because of its weak surface brightness and its location, the 
determination of the electron temperature of this particular velocity 
component may suffer from the lack of short-spacing data.

\section{Discussion}

The peaks of blue-shifted H92\( \alpha \) emission lie exactly on the 
three bundles of filaments that intersect the Sickle, supporting the 
notion that there is an interaction between the NTFs and the thermal 
ionized gas in the Sickle.  The results of the observations are 
generally consistent with the idea that cloud material ionized by UV 
radiation from hot stars is accelerated from the surface and some of 
the gas becomes attached to the magnetic filaments and is accelerated 
to relativistic velocities where it emits synchrotron radiation 
(Serabyn \& Morris 1994; Timmermann et al 1996).  Indeed, the 
synchrotron-emitting vertical filaments crossing the Sickle show an 
intrinsic positive spectral index (Yusef-Zadeh 1989; 
Anantharamaiah et al 1991; Tsuboi et al 1996), which becomes flat or 
negative toward more negative latitudes away from the Sickle (Pohl, 
Reich \& Schlickeiser 1992), whereas another group of non-thermal 
filamentary structures (the ``threads'') in the Galactic center 
region shows steep spectral indices but they are relatively isolated 
and do not coincide with any ionized thermal features (Anantharamaiah 
et al 1991; Gray et al 1994).  The flat spectrum of the Arc may 
indicate that thermal gas associated with the Sickle is mixed with 
synchrotron emitting nonthermal gas associated with the linear 
filaments (Yusef-Zadeh \& Morris 1987a; Anantharamaiah et al 1991).

Lang et al (1997) and Timmermann et al (1996) found a velocity 
gradient across the Sickle in a direction parallel to the filaments, 
with the largest (redshifted) velocities towards the west, and the 
smallest (roughly 20 \kms ) to the east.  The blue shifted gas 
generally lies to the east of the Sickle, consistent with this 
velocity gradient.  The densest ionized gas is dynamically coupled to 
the +25 \kms\ molecular cloud whereas the low-density diffuse features 
presumably originated at the outer surface of the cloud where they 
were accelerated to anomalous velocities.  Lang et al (1997) find an 
unusually high H115\( \beta \)/H92\( \alpha \) ratio, inconsistent 
with LTE within the Sickle where the NTFs intersect the Sickle.  The 
infrared observations of Simpson et al (1997) show that the extinction 
towards the Sickle region is uniform, implying that the Sickle itself 
lies on the front of the +25 kms\ cloud, and therefore that the 
ionised gas is being accelerated towards us from the surface of the 
cloud.  If the blue-shifted gas is being accelerated along the 
magnetic field aligned with the NTFs, then some of the filaments must 
be tilted so that the southeast points toward us.

If the magnetic field is responsible for the acceleration, the magnetic
pressure must dominate the ram pressure of the blue-shifted gas.  Adopting
an electron density of $\approx$150 cm$^{-3}$ for the $-35$ \kms feature at
position 1, a field strength of at least 0.2 mG is required.

Alternatively, the ram pressure from the combined winds of the hot 
mass-losing stars discovered near the Sickle (Moneti et al 1991; Figer 
et al 1995, 1996; Cotera et al 1996) may be sufficient to accelerate 
the gas.  The stars' broad He I and Br$\gamma$ emission lines indicate 
terminal velocities in the range $V_{w}=500-700$ km s$^{-1}$, similar 
to the cluster of young stars at the Galactic center (Krabbe et al 
1991).  Adopting a distance of 5 pc from the cluster to the Sickle, 
the required mass loss from hot stars with $V_{w}=700$ km s$^{-1}$ to 
accelerate the 30 \( M_\odot \) of ionised gas responsible for the 
feature at peak 1 from the surface of the Sickle at +25\kms\ to a 
velocity of -35\kms\ over a distance of 2pc is $\approx 3\times 
10^{-3}\;\,M_\odot$ yr$^{-1}$.

A ram pressure of this magnitude is also capable of accelerating the 
redshifted gas, for which the the physical relationship with the NTFs 
is less clear.  The gas at position 2 lies well away from the Sickle 
HII region, between the two northernmost bundles of filaments that 
pass through the Sickle.  A small continuum feature is present at this 
position in the 5 GHz image of Yusef-Zadeh \& Morris (1987a) and the 8.3 
GHz image of Lang et al (1997).  The southeastern end of the Sickle 
(near position 3) also does not appear to be asociated with any NTFs.  
The kinetic energy of the features are roughly \( 3 \times 10^{48} \) 
and \( 10^{48} \) erg respectively.

Measurements of the field strength and the total stellar mass loss rate in 
this region would be useful in discriminating between these 
mechanisms for accelerating the anomalous-moving clouds.

\acknowledgments

F. Yusef-Zadeh's work was supported in part by NASA grant NAGW-2518.
D. Roberts acknowledges support from the NSF grant AST94-19227. The 
SRCfTA is funded by the Australian Research Council under the Special 
Research Centres programme.

\section*{Figure Captions}

\noindent
Fig. 1. -- The gray scale image in the bottom right corner shows the 
8.3 GHz continuum emission from the Sickle region with a spatial 
resolution of 27\dasec8 $\times$ 24\dasec9 (PA=61$^\circ$).  Contour 
levels for this figure as well as Figs.  2, 3, and 4 are represented 
at 20, 80, 140, 200, and 260 mJy beam$^{-1}$.  The three 
H92$\alpha$ line profiles are obtained at the positions marked in the 
continuum image.  In the profiles, the crosses show the observed 
spectra, the solid lines show the model fits, and the dotted lines 
show the residuals.

\noindent
Fig. 2. -- The H92$\alpha$ line intensity integrated between $-63$ 
and $-7$ \kms\ is represented as gray scale with superposed contours 
of continuum emission at 8.3 GHz.  The peak emission from the 
blue-shifted velocity feature coincides at the location where the NTFs 
cross the Sickle.

\noindent
Fig. 3. -- The H92$\alpha$ line intensity integrated between +106 and 
+205 \kms\ is represented as gray scale with superposed contours of 
continuum emission at 8.3 GHz.

\noindent
Fig. 4. -- A pseudo-color representation of the velocity distribution 
of ionized gas covering the entire range between $-100$ and +190 \kms.  
The overlaid contours show the 8.3 GHz continuum emission in the 
Sickle region.

\end{document}